# Important role of magnetization precession angle measurement in inverse spin Hall effect induced by spin pumping

Surbhi Gupta[1,a)], Rohit Medwal[1], Daichi Kodama[1], Kouta Kondou[2], YoshiChika Otani[2,3], Yasuhiro Fukuma[1,2,a)]

[1]Department of Computer Science and Electronics, Kyushu Institute of Technology, Iizuka, Fukuoka 820-8502, Japan

[2] Center for Emergent Matter Science, RIKEN, 2-1 Hirosawa, Wako 351-0198, Japan

[3]Institute for Solid State Physics, University of Tokyo, Kashiwa 277-8581, Japan

**Abstract**

Here, we investigate spin Hall angle of Pt in $Ni_{80}Fe_{20}$/Pt bilayer system by using a broadband spin pumping and inverse spin Hall effect measurement. An out-of-plane excitation geometry with application of external magnetic field perpendicular to the charge current direction is utilized in order to suppress unwanted galvanomagnetic effects. Magnetization precession angle ($\theta_c$) on ferromagnetic resonance for wide excitation frequency range (4-14 GHz) is estimated from the rectification voltage of anisotropic magnetoresistance (AMR) and a conventional method of using microwave power in a coplanar waveguide. A marked difference in $\theta_c$ profiles for the different methods is observed, resulting in the large variation in estimated values of spin current density at $Ni_{80}Fe_{20}$/Pt interface. The frequency dependence of the spin current density estimated using AMR effect is found to be similar to that of the inverse spin Hall voltage. We obtain the frequency-invariant spin Hall angle of $0.067 \pm 0.002$.

---
a) Electronic mail: sgupta@fukuma-lab.info, fukuma@cse.kyutech.ac.jp

Over the past decades large number of experimental as well as theoretical attempts have been made to construct new insights and applications of spin pumping (SP) effect since Tserkovnyak, *et al.* theoretically proposed the SP effect to inject pure spin current ($J_S$) from a precessing ferromagnetic metal (FM) to neighboring normal metal (NM).[1] Indeed, SP effect driven by its versatility is being widely explored to determine spin to charge current conversion efficiency, referred as spin Hall angle ($\Theta_{SH}$), in various kind of materials ranging from strong spin-orbit coupling system of heavy metals[2-3], conventional semiconductors like (Si and Ge)[4-5], wide-band gap semiconductor oxides (ZnO, ITO)[6-7] to ferromagnetic metallic alloys[3,8] oxides like $SrRuO_3$[9] and even organic polymers[10]. In practice, other techniques of non-local injection in lateral spin valve structure[11], spin-torque ferromagnetic resonance[12-14] and time-resolved magneto-optical Kerr effect[15] have also been established by various groups to measure $\Theta_{SH}$ of different materials. However very recent reports of significant modification in SP efficiency by further introducing complex interfacial effects[16-17] and novel application of SP effect to probe magnetic phase transitions in antiferromagnetic systems[18-19] undoubtedly highlight that it is a subject of intensive research with enormous possibilities.

In a prototypical setup of spin pumping-inverse spin Hall effect (SP-ISHE) measurement, SP occurs during the excitation of ferromagnetic resonance (FMR) when precession of magnetization in FM injects pure spin current into adjacent NM layer and then the injected spin current is converted into the transverse dc voltage by means of ISHE in NM. Generally in SP-ISHE measurement set-up, other Galvanomagnetic effects such as anisotropic magnetoresistance (AMR), planar Hall effect and anomalous Hall effect also generate spurious dc voltage signal because of capacitive and/or inductive coupled microwave current in the FM/NM sample.[20] Most of the previously reported studies assumed that purely Lorentz line shape of the voltage signal originates from ISHE contribution whereas antisymmetric line shape corresponds to that from the unwanted rectification signals. However, this assumption holds true only for specially designed device configuration and for particular microwave frequencies, as clearly demonstrated by Harder et al.[21] and Obstbaum et al.[22] Therefore, systematic and controlled studies of line shape analysis in different external magnetic field direction were performed in order to carefully distinguish ISHE induced dc voltage ($V_{ISHE}$) unambiguously from unwanted spurious signal.[22-26]

However, in such studies, precession of the magnetization at FMR, i.e., the magnetization precession angle ($\theta_C$) and its trajectory [26-27] is not sufficiently characterized. It is important to estimate independently the precession angle ($\theta_C$) to have quantitative analysis of $\Theta_{SH}$ as $\Theta_{SH} \propto V_{ISHE} / J_S$, where $J_S$ is being second order effect in the precession angle ($\sin^2\theta_C$). Moreover, waveguide based experimental set-up shows different microwave power losses and in-turn varying $\theta_C$ values in the range of sweeping frequencies. Therefore these parameters should be characterized in such measurement scheme separately. Otherwise large disparity may happen in estimated values, which had already led to large inconsistency in $\Theta_{SH}$ value of the most studied metal Pt varying from 0.0067 to 0.11 and dubious frequency dependent behavior in spite of employing same method of the SP-ISHE measurement.[28] In this letter, we demonstrate the precise estimation of $\Theta_{SH}$ by independently evaluation of $\theta_C$ in each input microwave frequency. In order to determine the amplitude of $J_S$ caused by SP, the different evaluation methods, namely microwave magnetic field ($h_{rf}$) evaluation using input setting power ($P_{input}$)[29-30], transmitted power ($P_{trans}$)[31] and absorbed power ($P_{ab}$)[7,32] measurements and a direct evaluation of $\theta_C$ by using rectified dc voltage due to AMR effect[33-34], are compared to quantify frequency dependent $\theta_C$ in broad range of microwave frequency 4-14 GHz. In addition, out-of-plane microwave excitation geometry is opted for both $\theta_C$ and SP-ISHE measurements in the Py/Pt bilayer to completely avoid spurious rectification voltage signals. We observed similar behavior of $J_S$ estimated from the rectification voltage due to AMR and $V_{ISHE}$ as a function of applied microwave frequency, which provided reliable and invariant value of $\Theta_{SH}$ in the frequency range. On the other hand, $\Theta_{SH}$ deduced from $J_S$ determined from the $P_{input}$, $P_{trans}$ and $P_{ab}$ measurements showed significant variation.

Figure 1(a) and 1(b) show optical top-view images of two distinct devices designed to measure $\theta_C$ using AMR induced rectification voltage in Py (10 nm) layer and $V_{ISHE}$ in Py (10 nm)/Pt (10 nm) bilayer, respectively. In both devices, micro-strip with lateral dimensions of 300 × 5 μm$^2$ is fabricated using photolithography process. Coplanar waveguide (CPW) structure of Ti (5 nm)/Au(200 nm) was deposited such a way that Py and Py/Pt micro-strip lied in the space between ground and central signal line for inducing FMR in Py with out-of-plane microwave field ($h_{rf}$) as shown in Fig. 1(c). In addition lock-in-technique was employed to improve signal-to-noise ratio and therefore amplitude modulated microwaves referenced at 79 Hz were applied to CPW using a signal generator. For $\theta_C$ measurement using AMR, an additional dc current ($I_{dc}$) ranging from 0.1 to 0.5 mA was applied (Fig. 1(a)). Then lock-in-amplifier picked up the voltage signal as a function of sweeping external magnetic field $H_{dc}$ which can be rotated in-plane ($0 \leq \beta \leq 360°$) where $\beta$ is defined as angle of $H_{dc}$ with respect to the micro-strip (Fig. 1(c)). In-plane $H_{dc}$ field was applied parallel to the Py micro-strip ($\beta = 0$) in the $\theta_C$ measurement because its geometry then became sensitive to the modulation driven by oscillating resistance of Py. During $V_{ISHE}$ measurement, $H_{dc}$ was applied perpendicular to the strip ($\beta = 90°$).

Figure 2(a) shows dc voltage spectra for $I_{dc} = 0.4$ mA as a function of $H_{dc}$ for different applied microwave frequency ranging from 4 to 14 GHz, in the step size of 1 GHz with P$_{input}$ of 15 dBm. We chose high microwave signal of 15 dBm to compensate the transmission losses at multiple stages of microwave cable, connectors and probe station in the present setup. The detected spectrum showed complex line shape signal around the resonance field ($H_r$) for all the applied frequency range. The amplitude of the spectra ($V_\theta$) showed an expected decrease with increasing frequency because a torque that pulls the magnetization back into equilibrium condition increases with increasing $H_{dc}$ and therefore $\theta_C$ decreases.[34] The inset of Fig. 2(a) shows $V_\theta$ measured as a function of input microwave power ($P_{input}$) ranging from 13 to 18 dBm at fixed frequency of $f$ = 7 GHz and $I_{dc}$ = 0.3 mA. As microwave field amplitude ($h_{rf}$) increases as a square root of applied power and $\theta_C$ increases linearly with $h_{rf}$ field, the linear scaling of $V_\theta$ with microwave power with intercept at origin (0,0) shown in the inset is found to be consistent with the theory, implying that the present experiment was performed in a linear excitation regime.[20] The frequency-dependent behavior of $\theta_C$ was estimated using following equation[3, 33-34],

$$\theta_C = \sin^{-1}(\text{sqrt}(V_\theta/I_{dc} \times \Delta R_{AMR})) , \qquad (1)$$

where $\Delta R_{AMR} = R_\parallel - R_\perp$ is the maximum change in dc magnetoresistance and $R_\parallel$ and $R_\perp$ are resistance of the Py micro-strip in the saturated state when $H_{dc}$ is applied parallel and transverse to the applied current direction, respectively. An AMR ratio ($\Delta R_{AMR}/R_\parallel$) of approximately 0.5% was determined for Py using four probe method. Figure 2(b) depicts $\theta_C$ in Py as a function of applied microwave frequency obtained from the AMR and microwave power measurements. The microwave field $h_{rf}$ was estimated using Ampere's Law, $h_{rf} = \frac{\mu_o \sqrt{P/Z}}{2\pi w} \ln\left(1 + \frac{w}{D}\right)$ where $w$ is the width (10 μm) of the signal line in CPW and $D$ is the space (5 μm) between the signal line and the Py micro strip, $P$ is the microwave power ($P_{input}$ and $P_{trans}$) and $Z$ is the characteristic impendence of 50 Ω. Thereafter, frequency dependent $\theta_C$ in the limit of a small precession angle was extracted using $\theta_C = h_{rf}/2\Delta H_{Py}$,[16] where $\Delta H$ defines the half width of FMR spectra at the half maximum intensity. We also calculated $\theta_C$ using absorbed power ($P_{ab}$) which is estimated from FMR measurement using VNA[7,32] while the transmitting power ($P_{trans}$) is recorded using a power meter. For better understanding, the detail procedure of $P_{ab}$ and $P_{trans}$ measurement is shown in Fig. S1 provided in the supplementary section and frequency dependent behavior of $P_{ab}$ and $P_{trans}$ is also shown in supplementary Fig. S2. A continuous decease in $\theta_C$ as a function of the applied frequency is observed, however, the decreasing behavior is quite different for the measurement methods, which in turn further affect the determination of injected $J_S$.

To determine $J_S$ and $\Theta_{SH}$, frequency dependent $V_{ISHE}$ as a function of $H_{dc}$ is measured across Py/Pt bilayer micro-strip keeping same input power $P_{input}$ of 15 dBm. To avoid spurious AMR signal ($V_{AMR}$) from $V_{ISHE}$, we strictly aligned $H_{dc}$ at $\beta = 90º$ or $270º$ with respect to micro-strip. Such configuration led to zero $V_{AMR}$ contribution while $V_{ISHE}$ reaches to maximum amplitude as $V_{AMR}$ is proportional to $\sin2\beta$ whereas $V_{ISHE}$ exhibits $\sin\beta$ dependency in the out-of-plane $h_{rf}$ excitation configuration.[22,35] Figure 3(a) shows a symmetric Lorentz line shaped $V_{ISHE}$ spectrum as a function of $H_{dc}$ for different excitation frequencies where its sign reversal with equal magnitude under $H_r$ inversion, implies its ISHE origin. In support, the inset in Fig. 3(a) provides magnified view of purely symmetric line shape of detected voltage signal for $f = 10$ GHz where solid symbols depict experimental data while solid line represents the Lorentz fit. Then injected $J_S$ into the Pt layer upon FMR excitation for different applied frequency was calculated using[20]

$$J_s = \frac{2e}{\hbar} \times \frac{\hbar\omega}{4\pi} g^{\uparrow\downarrow} \sin^2\theta_C \left[ 2\omega \frac{\left(\gamma M_S + \sqrt{(\gamma M_S)^2 + (2\omega)^2}\right)}{(\gamma M_S)^2 + (2\omega)^2} \right] \qquad (2)$$

where $\omega = 2\pi f$ is the angular frequency, $g^{\uparrow\downarrow}$ is the spin mixing conductance of the Py/Pt interface, $\gamma = g\mu_B/\hbar$ is the gyromagnetic ratio, $g$ is the Lande's factor, $\mu_B$ is the Bohr magnetron and $M_S$ is the saturation magnetization. Frequency independent $g^{\uparrow\downarrow} = 1.9 \times 10^{19}$ m$^{-2}$ of Py/Pt interface is estimated using $g^{\uparrow\downarrow} = \frac{\gamma}{\omega}\frac{4\pi M_s t_{PY}\delta}{g\mu_B}$,[20] where $\delta$ accounts for broadening in line width of the FMR spectra ($\delta = \Delta H_{Py/Pt} - \Delta H_{Py}$) due to loss in angular spin momentum in Py during SP and $M_S$ of 870 mT is determined by fitting the Kittel formula, $\omega = \gamma\mu_0\sqrt{H_r(H_r + 4\pi M_S)}$.[20] It is important to mention here that the frequency dependence of $\theta_C$ defined in Eq. (2) for Py/Pt is not assumed to be same as precession cone angle $\theta_C$ of Py; first we determined the amplitude of $h_{rf}$ in the sample by using $h_{rf} = 2\theta_C \times \Delta H_{Py}$. Next, frequency dependence of $\theta_C$ for Py/Pt in the SP measurement configuration with $\beta = 90°$ is deduced from $\theta_C = h_{rf}/2\Delta H_{Py/Pt}$ because of the same design of both samples for Py and Py/Pt, as can be seen in Fig. 1. Figure 3(b) shows the frequency dependence of $J_S$ estimated from four different $\theta_C$ approaches as discussed in the previous paragraph. The behavior estimated from AMR measurement shows initial increment in the magnitude up to the excitation frequency of 10 GHz, followed by a decrease till 14 GHz, which is found to be similar to the experimentally obtained trend of $V_{ISHE}$ as shown in Fig. 3(a). On the other hand, $J_S$ values estimated from the microwave power measurements showed monotonic decrease and/or increase with applied frequency. Note that for $f = 10$ GHz, $J_S \sim 5.8 \times 10^6$ A/m$^2$ is determined from $P_{trans}$ measurement which is roughly one order less than $J_S$ value of $4.4 \times 10^7$ A/m$^2$ obtained from AMR method. Whereas $J_S$ values estimated from $P_{input}$ as well as $P_{ab}$ showed uplift in whole frequency range due to the overestimated values of $\theta_C$ as shown in Fig. 2(b). Such a large variation in the estimated values of $J_S$ further resulted into the quite different values of $\Theta_{SH}$ for the same Py/Pt sample. As injected $J_S$ converts into dc transverse charge current $J_C$ via ISHE due to high spin orbit coupling in Pt, defined by $J_C = \Theta_{SH} \times J_S$, the resultant voltage $V_{ISHE}$ generated along the Py/Pt sample can be written as[20]

$$V_{\text{ISHE}} = \Theta_{\text{SH}} J_S \left[ \frac{L}{\sigma_{Pt} t_{Pt} + \sigma_{Py} t_{Py}} \lambda_{sd} \tanh\left(\frac{t_{Pt}}{2\lambda_{sd}}\right) \right] \quad (3)$$

where $L$ is the length of the Py/Pt strip, $\sigma_{Pt}$ and $\sigma_{Py}$ are conductivities of Pt and Py layer, $t_{Pt}$ and $t_{Py}$ are their respective thicknesses and $\lambda_{sd}$ is the spin diffusion length of Pt. Considering $\lambda_{sd}$ = 1.2 nm as previously reported for the sample fabricated in the same conditions[13] and $\sigma_{Pt}$ = 2.5 × 10$^6$ $\Omega^{-1}$m$^{-1}$ and $\sigma_{Py}$ = 2.86 × 10$^6$ $\Omega^{-1}$m$^{-1}$ as experimentally measured for present samples, the estimated $\Theta_{\text{SH}}$ as a function of applied frequency (4-14 GHz) is plotted in Fig. 4. $\Theta_{\text{SH}} \sim 0.067 \pm 0.002$ of Pt using independent measurement of $\theta_C$ from the AMR method revealed a frequency invariant behavior while $\Theta_{\text{SH}}$ values determined from $P_{\text{trans}}$ showed a continuous and steep increase from 0.02 to 0.33 in the given microwave frequency range. To further emphasize the $\theta_C$ role, we also plotted $\Theta_{\text{SH}}$ values determined using $P_{\text{input}}$ in the inset of Fig. 4 where underestimated values of $\Theta_{\text{SH}}$ also showed frequency dependent behavior. Important to mention here that smaller $\Theta_{\text{SH}}$ values determined from $P_{\text{ab}}$ measurement also showed a steady behavior in the frequency range, however, the monotonic increase in $J_S$ values with the microwave frequency in this case was also not consistent with $V_{\text{ISHE}}$. Therefore the experimental data presented above clearly demonstrate that independent AMR measurement of $\theta_C$ is important, especially for waveguide setup. The utilization of out-of-plane excitation geometry in the Py strip helps into the unambiguous determination of symmetric Lorentzian line shape $V_{\text{ISHE}}$ signal. This combined methodology is necessary to obtain frequency invariant $\Theta_{\text{SH}}$ which is a material specific parameter.

In summary, we investigated the indispensable role of independent estimation of $\theta_C$ in the analysis of SP-ISHE effect, specifically in case of integrated co-planar waveguide architecture. Direct estimation of $\theta_C$ using rectified dc voltage due to AMR showed relatively small change in $\theta_C$ magnitude with an increase of applied microwave frequency whereas frequency dependent $\theta_C$ behavior estimated from conventional method of employing microwave power showed a monotonic decrease. As a result, estimated values of $J_S$ differed by more than an order of magnitude. Most important, the frequency dependent $J_S$ behavior estimated using the AMR measurement is found to be similar trend to experimentally obtained values of $V_{ISHE}$ amplitude, resulted in the frequency-invariant spin Hall angle ($\Theta_{SH} \sim 0.067 \pm 0.002$) of Pt for the studied frequency interval of 4-14 GHz. This approach essentially offers a simplified and reliable way to investigate the fundamental parameters of spin Hall effect in new promising materials using SP-ISHE measurements.

**Supplementary Material**

See supplementary material for complete understanding of the detail procedure of microwave power ($P_{ab}$) and ($P_{trans}$) measurement.

**Acknowledgement**

Financial support from Grant-in-Aid for Encouragement of Young Scientists (Grant No. 16K18079) and Grant-in-Aid for Scientific Research on Innovative Area, "Nano Spin Conversion Science" (Grant No. 26103002) are gratefully acknowledged. This work was partially supported by the Asahi Glass Foundation.

**Figure Caption**

**Fig. 1.** Schematic of experimental setup and optical image of respected devices for measurement of (a) AMR voltage, (b) ISHE voltage, and (c) enlarged view of Py strip, where $\beta$ defines direction of in-plane $H_{dc}$ with respect to central line.

**Fig. 2.** (a) Dc voltages as a function of $H_{dc}$ at $I_{dc} = 0.4$ mA for different frequency of 4-14 GHz (in step size of 1 GHz) at excitation power of 15 dBm. Inset shows linear scaling of voltage with input power (13-18 dBm) (b) Comparative plot of precession angle versus frequency determined by AMR-method and $h_{rf}$ evaluation from microwave power ($P_{trans}$, $P_{input}$ and $P_{ab}$).

**Fig 3.** (a) Frequency dependent spectrum of $V_{ISHE}$ as a function of $H_{dc}$ for Pt/Py in step of 1 GHz. Inset shows $V_{ISHE}$ signal at $f = 10$ GHz, solid circles are experimental data and line is Lorentz fit (b) Frequency dependent spin current density ($J_S$) obtained using AMR measurement and $h_{rf}$ evaluation from microwave power ($P_{trans}$, $P_{input}$ and $P_{ab}$).

**Fig.4.** Frequency dependent $\Theta_{SH}$ values obtained by AMR measurement and $h_{rf}$ evaluation from microwave power ($P_{trans}$, $P_{input}$ and $P_{ab}$). The dashed line represent average $\Theta_{SH}$ value.

Fig. 1.

Fig. 2.

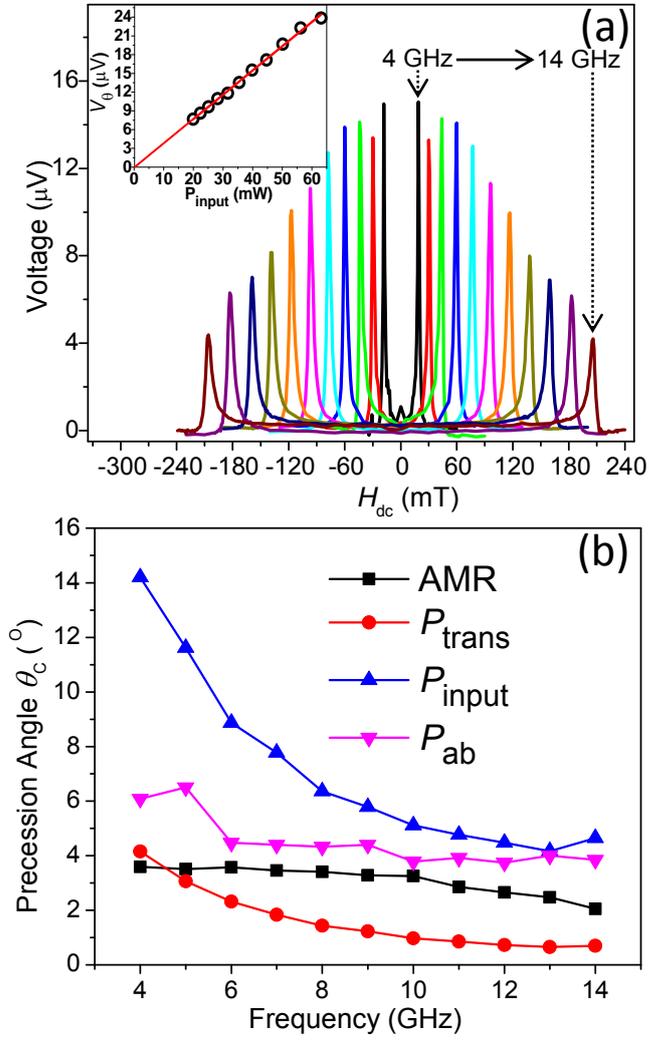

Fig. 3.

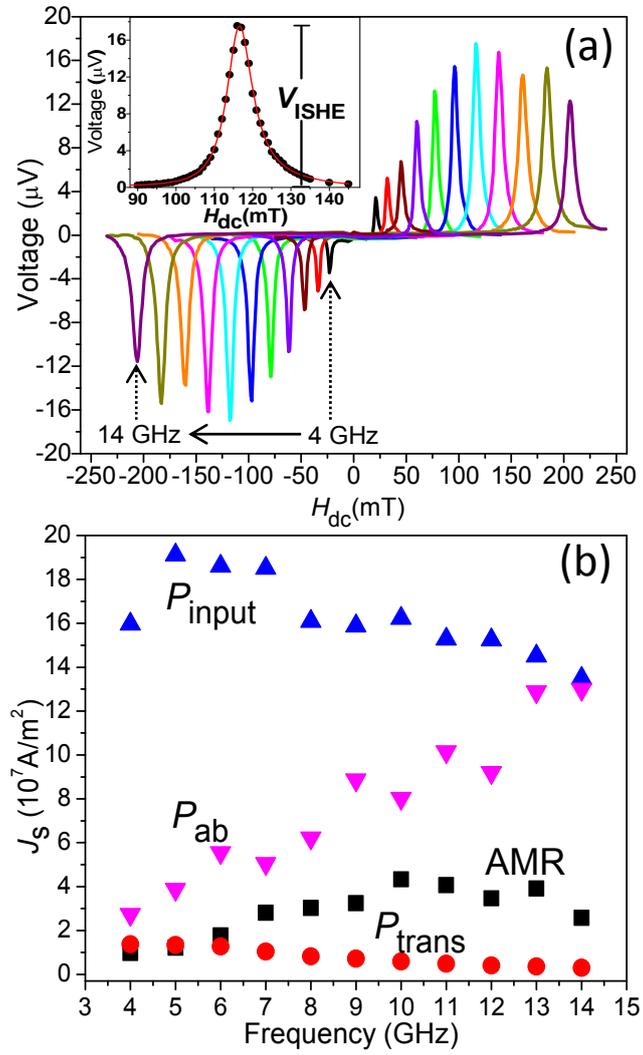

Fig. 4.

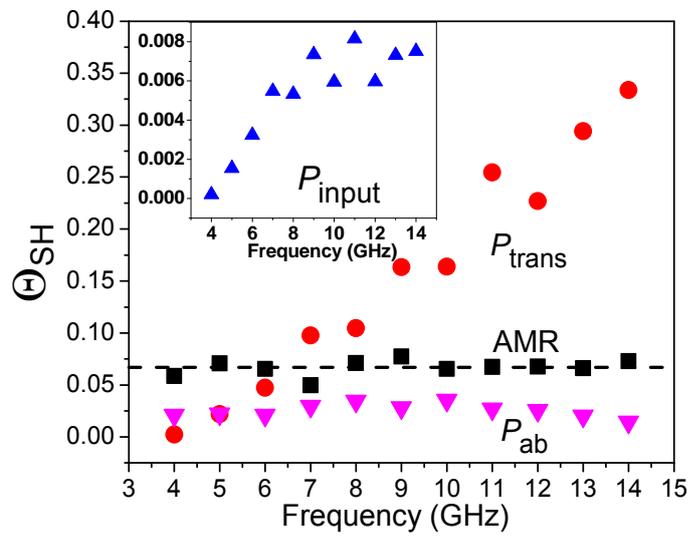

Supplementary Material

Microwave transmission intensity ($P_{trans}$) was measured using power meter (Agilent 53152A) for different applied frequencies, as shown in Fig. S1 (a). The magnetic field $h_{rf}$ applied to the sample was estimated using Ampere's Law, $h_{rf} = \frac{\mu_o \sqrt{P_{trans}/Z}}{2\pi w} \ln\left(1 + \frac{w}{D}\right)$ where $w$ is the width (10 μm) of the signal line in CPW and $D$ is the space (5 μm) between the signal line and the Py micro strip and $Z$ is the characteristic impendence of 50 Ω. On the other hand, microwave absorption intensity ($P_{abs}$) is estimated from the peak amplitude in the Lorentzian spectra of two-port microwave transmission property $|S_{21}|^2$ measured by the vector network analyzer (Agilent N5222A) as shown in Fig. S1(b).

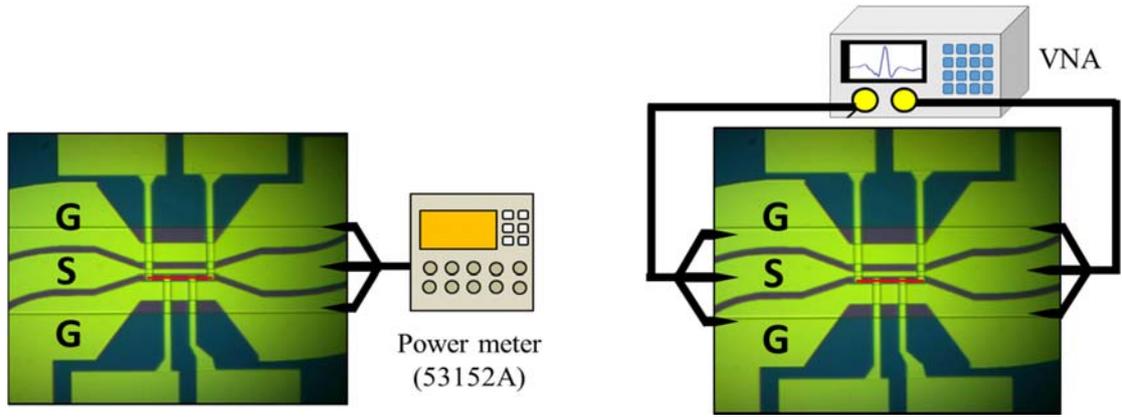

Fig. S1. Schematic for measurement of (a) transmission power $P_{trans}$ and (b) absorption power $P_{ab}$.

Here $|S_{21}|^2$ denotes the ratio of transmitted microwave power at port 2 to the incident fixed microwave power $P_{input}$ at port 1.[7, 26] Frequency dependent $h_{rf}$ distribution using $P_{ab}$ can be described as

$$P_{ab} = \upsilon \times \frac{\mu_o \gamma M_S}{4\alpha} h_{rf}^2 \left[\frac{\left(\gamma M_S + \sqrt{(\gamma M_S)^2 + (2\omega)^2}\right)}{(\gamma M_S)^2 + (2\omega)^2}\right]$$

where υ denotes the volume of the region that is magnetically excited by the microwave in the Py layer and α is the damping constant. Figure S2 depicts the frequency dependent $P_{ab}$ and $P_{trans}$ values which showed different behavior and as a result different $\Theta_{SH}$ values are obtained. It is important to mention here that accurate estimation of $J_S$ as well as $\Theta_{SH}$ requires accurate value of $\theta_C$. In earlier reports, $P_{input}$ is utilized to determine $h_{rf}$ without taking care of the actual transmission losses in CPW, which leads to overestimated values of $\theta_C$ as shown in the present study. Considering frequency dependent $P_{ab}$ and $P_{trans}$ measurements, we observed different behavior than AMR based results, suggesting that the indirect $\theta_C$ measurement techniques provide only a tentative trend.

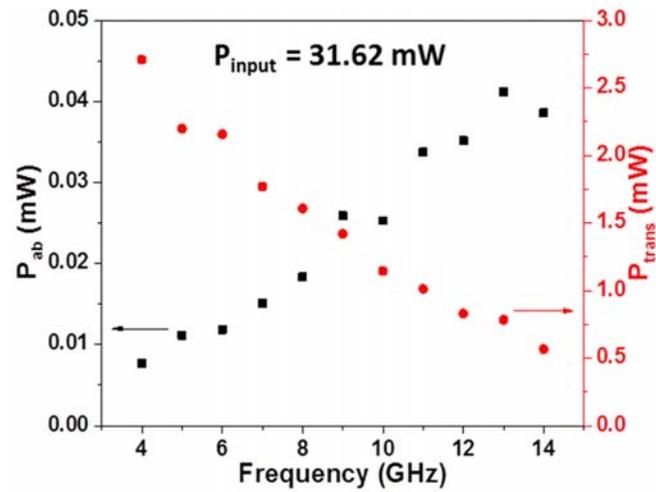

Fig. S2. Frequency dependent of $P_{ab}$ and $P_{trans}$ measured for Py sample